\documentstyle[twocolumn,prl,aps,psfig]{revtex}

\begin{document}

\draft \flushbottom \twocolumn[
\hsize\textwidth\columnwidth\hsize\csname @twocolumnfalse\endcsname

\title{Influence of oxygen isotope exchange on the ground state of manganites}
\author{V. N. Smolyaninova$^1$, Amlan Biswas$^1$, P. Fournier$^2$, S. Lofland$^3$,
 X. Zhang$^1$, Guo-meng~Zhao$^4$, and
 R. L. Greene$^1$}
\address{1. Department of Physics and Center for Superconductivity Research,\\
University of Maryland, College Park, MD 20742}
\address{2. Centre de recherche sur les propri\'et\'es \'electroniques de mat\'eriaux
avanc\'es and D\'epartement de Physique, Universit\'e de
Sherbrooke, Sherbrooke, Qu\'ebec, CANADA, J1K 2R1}
\address{3. Department of Chemistry and Physics, Rowan University,
Glassboro, NJ 08028-1701}
\address{4. Texas Center for Superconductivity, University of Houston,
Houston, TX 77204-5932}
\date{\today}
\maketitle
\tightenlines
\widetext
\advance\leftskip by 57pt
\advance\rightskip by 57pt

\begin{abstract}
We report a study of oxygen isotope effects on low temperature
specific heat, magnetization, and resistivity of
La$_{1-x}$Ca$_{x}$MnO$_{3}$ and
(La$_{1-y}$Pr$_{y}$)$_{0.67}$Ca$_{0.33}$MnO$_{3}$. For the
metallic compositions of La$_{1-x}$Ca$_{x}$MnO$_{3}$ and for
charge-ordered La$_{0.5}$Ca$_{0.5}$MnO$_{3}$ no change in low
temperature specific heat has been detected with $^{16}$O -
$^{18}$O exchange, while  compounds of
(La$_{1-y}$Pr$_{y}$)$_{0.67}$Ca$_{0.33}$MnO$_{3}$ ($0.4<y<0.6$)
show a significant change in low temperature properties. The
low temperature specific heat indicates a presence of the
charge-ordered phase even in compositions of
(La$_{1-y}$Pr$_{y}$)$_{0.67}$Ca$_{0.33}$MnO$_{3}$ which are
metallic at low temperatures. We suggest that the changes induced
by the oxygen isotope exchange are caused by an increase of the
charge-ordered phase in $^{18}$O samples.

\vspace{0.5cm}

PACS number(s): 75.40.Cx, 75.30.Vn, 71.30.+h, 75.50.Cc
\end{abstract}

]
\narrowtext
\tightenlines

\section{INTRODUCTION}
Manganese oxides of general formula R$_{1-x}$A$_{x}$MnO$_{3}$
(where R is a rare-earth ion and A is an alkaline-earth ion) have
attracted considerable interest due to their various fascinating
properties. They exhibit a rich  phase diagram, which includes
ferromagnetic (FM) metallic, antiferromagnetic (AFM) insulating,
charge and orbitally ordered phases depending on doping $x$,
average ionic radius $<r_{\rm A}>$ on R and A sites, magnetic
field, and temperature. In manganites, charge, spin, orbital, and
lattice degrees of freedom are interconnected, and their balance
determines the electronic and magnetic state of these materials.
In some compositions a competition of opposing interactions
associated with these different degrees of freedom induces an
electronic phase separation, which is considered to be an
intrinsic properties of these manganites \cite{Uehara,fath}.
Another prominent manifestation of the spin-charge-lattice
interplay in manganites is the unusual and large oxygen isotope
effect \cite{GMNature,TCO,MI}. Substitution of $^{16}$O by
$^{18}$O lowers significantly the Curie temperature $T_{\rm C}$ of
the ferromagnetic compositions of La$_{1-x}$Ca$_{x}$MnO$_{3}$
\cite{GMNature,Franck1}, increases the charge ordering (CO)
transition temperature $T_{\rm CO}$ for the transitions from FM
metallic to AFM charge-ordered state \cite{TCO} and even induces a
metal-insulator transition \cite{MI}. Many models have been
proposed to explain the isotope effect in manganites: a small
polaron model \cite{GMNature,Yu}, a bipolaron model \cite{GMbi},
change of interatomic distance by lattice vibrations
\cite{Belova1}, nonadiabatic behavior of the oxygen ions
\cite{Kresin}, isotope dependence of the nonstoichiometry in
manganites  \cite{Nagaev}. In spite of significant experimental
and theoretical effort, a physical picture that would consistently
explain the influence of oxygen isotope substitution on different
states (FM and CO) is presently lacking. To better understand the
nature of the oxygen isotope effect in manganites, it is important
to know how it affects the ground state. Studies of low
temperature specific heat can address this issue, since the
specific heat carries information about principal excitations.

In this paper we report a systematic study of specific heat,
magnetization and resistivity in FM metallic composition of
La$_{1-x}$Ca$_{x}$MnO$_{3}$ ($x = 0.2$, 0.3, and 0.375),
charge-ordered La$_{0.5}$Ca$_{0.5}$MnO$_{3}$, as well as
(La$_{1-y}$Pr$_y$)$_{0.67}$Ca$_{0.33}$MnO$_{3}$ ($0<y<1$). The
latter compound changes from FM metal for $y=0$ to CO AFM
insulator for $y=1$ \cite{Cheong1}. For the compositions  $y<0.75$
the (La$_{1-y}$Pr$_y$)$_{0.67}$Ca$_{0.33}$MnO$_{3}$ exhibits a
percolative transition from a higher temperature CO insulator to a
metal with decreasing temperature \cite{Uehara,Cheong1,Podzorov}.
Analysis of low temperature magnetization, resistivity
\cite{Cheong1,Belova2}, and synchrotron x-ray diffraction data
\cite{Kiryukhin} indicates metallic and CO phase coexistence in
the $y<0.75$ compositions of these materials. One goal of this
paper is to  compare the low temperature behavior of these various
samples in order to investigate the influence of oxygen isotope
exchange on the properties of both single phase and phase
separated  manganites. We have found that the most prominent
isotope effect is observed in the phase separated systems, which
might be the key to understanding the origin of the giant oxygen
isotope effect in manganites.

\section{EXPERIMENTAL DETAILS}

Ceramic samples of La$_{1-x}$Ca$_{x}$MnO$_{3}$ ($x = 0.2$, 0.3, 0.375, and 0.5)
and (La$_{1-y}$Pr$_y$)$_{0.67}$Ca$_{0.33}$MnO$_{3}$ ($0<y<1$)
were prepared by a standard solid state reaction technique. X-ray powder
diffraction showed that all samples are single phase and good quality.
For oxygen isotope exchange we used a standard procedure \cite{GMNature}:
the $^{16}$O and  $^{18}$O samples were prepared from the same pellet of the
starting material, and were simultaneously treated at T = 1000$^{\circ}$ C and pressure
1 atm in different closed quartz tubes, one filled with $^{16}$O and the other
with  $^{18}$O. We estimate the extent of the exchange of $^{16}$O by $^{18}$O to be at
least 85 \%, as determined by weight change.

The specific heat was
measured in the temperature range 2-20 K and magnetic field range 0-8.5~T by
relaxation calorimetry.
In the temperature range 20-300 K the specific
heat was measured by a Quantum Design PPMS. The specific heat measurements have
an absolute accuracy of $\pm 3$\%. The magnetization was measured with a
commercial SQUID magnetometer in magnetic field range 0-5.5 T and  with
a Quantum Design PPMS in fields up to 9 T. Resistivity was measured by a standard four-probe
technique.

\section{RESULTS}

\subsection{Low temperature specific heat of La$_{1-x}$Ca$_{x}$MnO$_{3}$ with $^{16}$O and
$^{18}$O  }

In our samples, substitution of $^{16}$O by $^{18}$O decreases the
Curie temperature (determined from the temperature dependence of
the magnetization in a magnetic field of 50 G) of
La$_{1-x}$Ca$_{x}$MnO$_{3}$ significantly: T$_C$($^{16}$O) -
T$_C$($^{18}$O) = 17.7 $\pm 0.5$ K, 11 $\pm 0.5$ K, 6 $\pm 0.5$ K
for $x = 0.2$, 0.3, and 0.375 respectively. The $x$ dependence of
the oxygen isotope effect and its values are consistent with prior
work \cite{GMNature,Franck1,GM2}. Our low temperature specific
heat measurement of the $^{16}$O and $^{18}$O  samples of these
compositions  is the same within the experimental error ($\pm
2$\%) in the temperature range from 2 K to 10 K, as shown in Fig.
1a,b for $x = 0.2$ and 0.3. Since these materials are in the FM
metallic state at low temperatures we fit the $x = 0.2$ and 0.3
data    to a form $C=\gamma T+\beta T^{3}+\delta T^{3/2}$ for
$4.2<T<8$ K, where $\gamma T$ is the charge carrier contribution,
$\beta T^3$ the phonon contribution, and $\delta T^{3/2}$ the
ferromagnetic spin-wave contribution. This fitting range  was
chosen, because for $T<4$~K the hyperfine contribution from the
nuclear magnetic levels of Mn ions may appear, and for $T>\Theta
_{\rm D}/50$, where $\Theta _{\rm D}$ is the Debye temperature,
the higher terms of the lattice expansion ($\beta _{5}T^{5}$,
$\beta _{7}T^{7}$ etc.) could be present. We found that values of
the respective fitting parameters for $^{16}$O and $^{18}$O
samples ($\gamma _{16}=5.7 \pm 0.1$ mJ/mole K$^2$,
$\gamma_{18}=5.6 \pm 0.1$ mJ/mole~K$^2$, $\beta _{16}=0.178 \pm
0.003$ mJ/mole K$^4$, and $\beta _{18}=0.177 \pm 0.003$ mJ/mole
K$^4$ for $x=0.2$ samples and $\gamma _{16}=5.8 \pm 0.1$ mJ/mole
K$^2$, $\gamma_{18}=5.8 \pm 0.1$ mJ/mole K$^2$, $\beta _{16}=0.180
\pm 0.003$~mJ/mole K$^4$, and $\beta _{18}=0.177 \pm 0.003$
mJ/mole K$^2$ for $x=0.3$ samples) coincide within experimental
error. The best fit requires $\delta =0$. As was noted in previous
work \cite{Hamilton,sph}, it is difficult to resolve the
ferromagnetic spin-wave contribution to the specific heat in FM
metallic manganites due to its small value and the presence of the
$\gamma T$ contribution.
\begin{figure}[tbp]
\centerline{ \psfig{figure=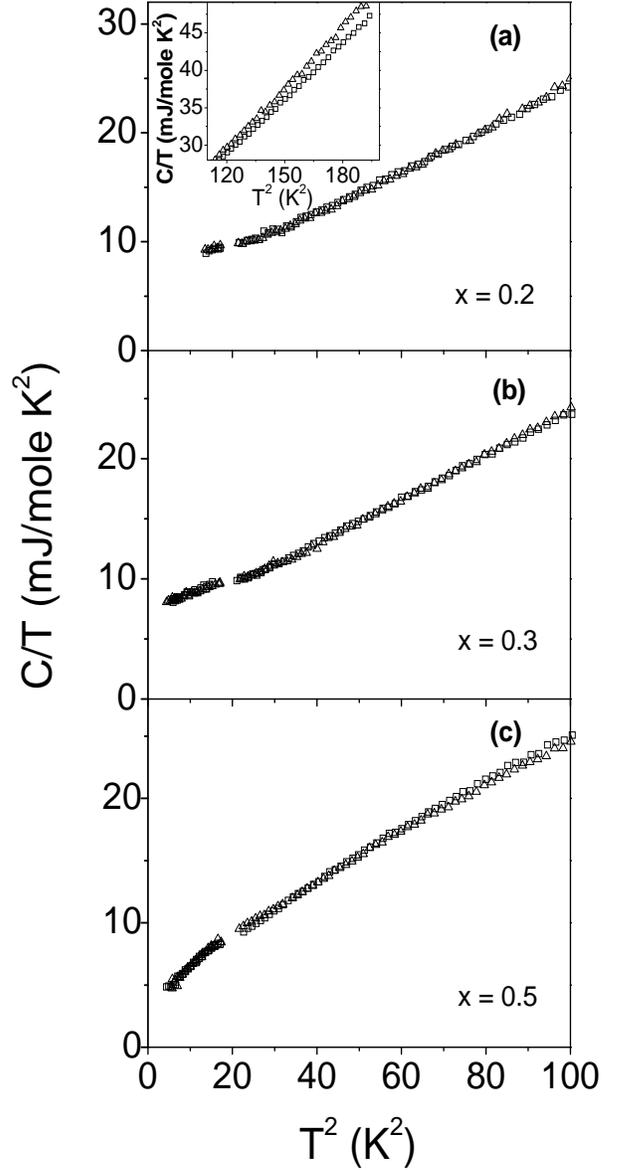,width=8.5cm,height=16.2cm} }
\caption{Specific heat of La$_{1-x}$Ca$_{x}$MnO$_{3}$ samples with
$^{16}$O (squares) and $^{18}$O (triangles)  plotted as $C/T$ vs
$T^{2}$ in temperature range from 2 to 10 K: (a) $x = 0.2$, (b) $x
= 0.3$, (c)  $x = 0.5$. Inset shows $C/T$ vs $T^{2}$ for the $x =
0.2$ sample in temperature range from 11 to 14 K. } \label{fig1}
\end{figure}
\noindent Although we cannot resolve the spin-wave contribution
from our data, it should not affect our conclusions, since
inelastic neutron scattering \cite{Lynn} shows that the spin
dynamics remains the same for $^{16}$O and $^{18}$O
La$_{1-x}$Ca$_{x}$MnO$_{3}$ in this composition range at low
temperatures. The  values of the charge carrier ($\gamma T$) and
lattice contribution ($\beta T^{3}$) are close to that found in
other FM metallic manganites \cite{Hamilton,sph}.

Although we did not observe a change in the  lattice contribution to the
specific heat in the temperature range $4.2<T<8$ K, the $^{18}$O samples should
have a larger lattice contribution, since the frequencies of the
lattice vibrations should be smaller for the heavier oxygen. We do not expect any
measurable contribution from optical modes in this temperature range.
For estimates of the changes in the lattice contribution with oxygen isotope
substitution we consider $\Theta _{\rm D} \propto M^{-1/2}$, and therefore
$\beta \propto M^{3/2}$. This change in $\beta$ corresponds to only a 2\% (below our
resolution limit) increase
of total specific heat for $T=6$ K, since the specific heat at low temperatures
is dominated by the charge carrier contribution. However, at higher temperatures,
where the specific heat is dominated by the lattice contribution, the specific
heat of the $^{18}$O sample is larger, than that of the $^{16}$O sample (Fig. 1a,
inset), which corresponds to a larger lattice contribution. The magnitude of this
increase  is consistent with estimates for the change in the lattice specific heat
described above.
After oxygen back-exchange ($^{16}$O $\rightarrow$ $^{18}$O;
$^{18}$O $\rightarrow$ $^{16}$O) the $^{16}$O -$^{18}$O pairs of samples showed
the same results:
a significant change in $T_{\rm C}$ and no change in the low
temperature specific heat within experimental error.

In the charge-ordered compound La$_{0.5}$Ca$_{0.5}$MnO$_{3}$ oxygen isotope substitution
strongly increases the
charge ordering temperature \cite{TCO}, which coincides  for this compound with the temperature
of the transition from FM to AFM state:
T$_{\rm CO}$($^{16}$O) - T$_{\rm CO}$($^{18}$O) = 11.7 $\pm 0.5$ K
(determined from the temperature dependence of the magnetization measured in $B=50$ G and the
temperature dependence of the resistivity),
but the low temperature specific heat remains unchanged (Fig. 1c).
Thus, for single phase metallic and CO compositions the specific heat does not
show any change in principal excitations (except for the expected increase of the lattice
contribution for $^{18}$O samples which we observed for $T>11$ K, as mentioned above), which
indicates that the ground state of
these materials remains unchanged after oxygen isotope substitution.

\subsection{Low temperature specific heat  of
 (La$_{1-y}$Pr$_y$)$_{0.67}$Ca$_{0.33}$MnO$_{3}$ ($0<y<1$)}

To study the influence of oxygen isotope substitution on the
materials with metallic and CO phase coexistence we chose
(La$_{1-y}$Pr$_y$)$_{0.67}$Ca$_{0.33}$MnO$_{3}$ ($0<y<1$). Before
presenting the experimental results for $^{16}$O - $^{18}$O
samples, it is important to understand how compositional ($y$)
variations affect the low temperature properties. Fig. 2 shows the
temperature dependence of the resistivity, $\rho (T)$, of our
(La$_{1-y}$Pr$_y$)$_{0.67}$Ca$_{0.33}$MnO$_{3}$ ($0<y<1$) samples.
The magnetization  values at $\mu _0 H = 1$ T and $T = 5$~K (after
zero field cooling (ZFC)) for different $y$ is shown in the inset
of Fig. 2. For the $y=0.4$, 0.5, and 0.6 compositions the
temperature dependence of the resistivity changes from insulating
at higher temperatures to metallic at low temperatures. The
magnetization  values at $\mu _0 H  = 1$ T (slightly above
saturation field, but below the field where the CO is affected by
the magnetic field) decreases with increase of $x$ indicating that
the fraction of the FM phase decreases for compositions richer in
Pr (Fig. 2, inset). This behavior of the resistivity and
magnetization is consistent with previous work \cite{Cheong1}.

\begin{figure}[tbp]
\centerline{
\psfig{figure=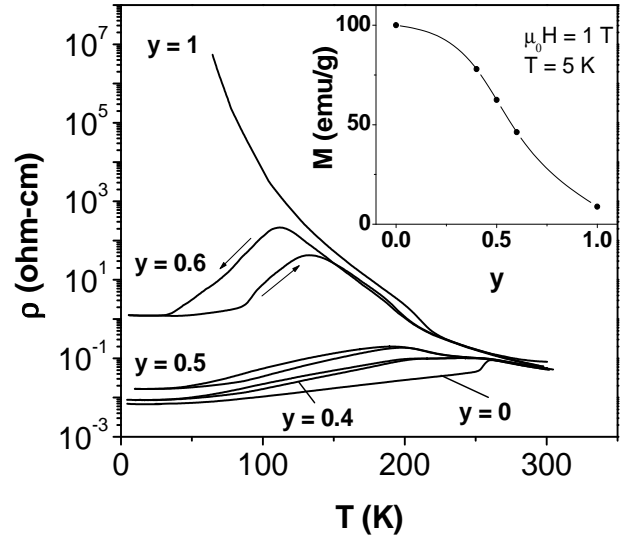,width=8.5cm,height=7.5cm,clip=} }
\caption{Temperature dependence of the resistivity of
(La$_{1-y}$Pr$_y$)$_{0.67}$Ca$_{0.33}$MnO$_{3}$ ($0<x<1$) on
warming and cooling. Inset shows ZFC magnetization at $T = 5$ K
and $\mu _0 H =1$~T for different~$y$. } \label{fig2}
\end{figure}

The low temperature specific heat  of (La$_{1-y}$Pr$_y$)$_{0.67}$Ca$_{0.33}$MnO$_{3}$
($0<y<1$) is shown in Fig. 3. The specific heat of La$_{0.7}$Ca$_{0.3}$MnO$_{3}$ in the
temperature range from 4 K to 19 K is given by
\begin{equation}
C = \gamma T + \beta T^3 + \beta _5 T^5
\end{equation}
where the higher term of the lattice contribution $\beta _5 T^5$ ($\beta _5 = 0.08 \pm 0.001$
$\mu$J/mole-K$^6$) is required to fit the data in this
temperature range. We express the specific heat of Pr$_{0.65}$Ca$_{0.35}$MnO$_{3}$ in the
temperature range from 4 K to 12 K in the following form \cite{Vera2}:
\begin{equation}
C = \gamma T + \beta T^3 + C^{\prime}(T)
\end{equation}
The first term in this electrically insulating sample originates from spin and charge
disorder \cite{Vera2}. The second term represents the lattice and the AFM spin-wave
contribution to the specific heat. The third term, $C^{\prime}(T)$, is an anomalous
contribution present only in the CO state \cite{Vera2,Vera1}, which manifests itself as
upward curvature in the $C/T$ vs. $T^2$ plot. All the samples at intermediate
compositions ($y=0.4$, 0.5, and 0.6)  exhibit the $C^{\prime}(T)$ term even
though the low temperature  resistivity of $y=0.4$
\begin{figure}[tbp]
\centerline{
\psfig{figure=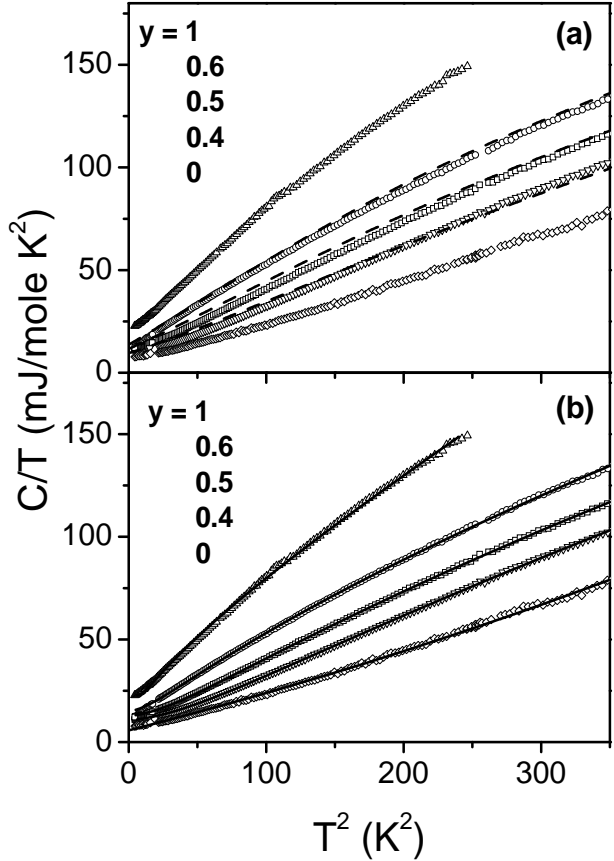,width=8.5cm,height=12cm,clip=} }
\caption{Specific heat of
(La$_{1-y}$Pr$_y$)$_{0.67}$Ca$_{0.33}$MnO$_{3}$: up triangles, $y=
1$; circles, $y=0.6$; squares, $y=0.5$; down  triangles, $y =
0.4$; diamonds, $y=0$. Dashed lines in (a) are described in text.
Solid lines in (b) are fits described in text. } \label{fig3}
\end{figure}
\noindent and 0.5 samples is similar to metallic
La$_{0.7}$Ca$_{0.3}$MnO$_{3}$. This indicates a presence of CO for
metallic (at low T) compositions of
(La$_{1-y}$Pr$_y$)$_{0.67}$Ca$_{0.33}$MnO$_{3}$ ($y=0.4$, 0.5, and
0.6) and, therefore, coexistence of CO insulating and
\begin{table}[tbp]
\begin{center}
\begin{tabular}{||c|c|c|c|c||}
y&$\gamma$&$\beta$&$\Delta$&$B$\\
((La$_{1-y}$Pr$_y$)$_{0.67}$Ca$_{0.33}$MnO$_{3}$)  &  &  &  &\\
\hline 1   &  15.7 & 0.39 & 1.15 & 20.7 \\ \hline 0.6 &  14.8 &
0.159 & 1.65 & 10.6 \\ \hline 0.5 &  12.5 & 0.117 & 2.21 & 8.28 \\
\hline 0.4 &  10.4 & 0.103 & 2.67 & 7.31 \\ \hline 0   &   5.8 &
0.18  &  &  \\ \hline 0.5 ($^{18}$O) & 14.9 & 0.183 & 1.51 & 10.8
\\ \hline
0.4 ($^{18}$O) & 12   & 0.166 & 2.23 & 11.1 \\ \hline 0.5
($^{18}$O $\mu _0 H  = 8.5$ T) & 9.6  & 0.13  &  2.31 & 7.79 \\
\hline 0.5 ($^{16}$O $\mu _0 H  = 8.5$ T) & 7.0  & 0.108 &  2.72 &
6.59 \\ \hline
\end{tabular}
\end{center}
\caption{Summary of the fitting results for the specific heat
data. The units of different quantities are:
$\gamma$~(mJ/mole-K$^2$),  $\beta$  (mJ/mole-K$^4$),  $\Delta$
(meV), and $B$  (meV-\AA$^2$).}
\end{table}
\noindent metallic phases. The presence of charge ordering below
the metal-insulator transition was also observed by synchrotron
x-ray diffraction in $y \approx 0.5$ \cite{Kiryukhin}.

The specific heat  of
(La$_{1-y}$Pr$_y$)$_{0.67}$Ca$_{0.33}$MnO$_{3}$ increases
consistently with increase of the Pr content $y$, suggesting the
increase of the volume fraction of the CO phase. To test this
assumption we consider  the fraction of the metallic phase in
(La$_{1-y}$Pr$_y$)$_{0.67}$Ca$_{0.33}$MnO$_{3}$ to be proportional
to the saturation magnetization  value
\cite{Cheong1}. Since the magnetization is almost temperature
independent in the temperature range from 2 K to 20 K, the
fractions of the FM metallic phase $f_{\rm met}=M(y)/M(0)$ in this
temperature region are $0.80 \pm 0.02$, $0.62 \pm 0.01$, and $0.45
\pm 0.01$ for $y=0.4$, 0.5, and 0.6 respectively (Fig. 2, inset).
If we suppose that the FM fraction has the specific heat of
La$_{0.7}$Ca$_{0.3}$MnO$_{3}$ ($C_{\rm met}$) and the CO fraction
has the specific heat of Pr$_{0.65}$Ca$_{0.35}$MnO$_{3}$ ($C_{\rm
CO}$), then the specific heat of such a two phase system  can be
expressed as $C_{\rm tp}(T)=f_{\rm met}C_{\rm met}+(1-f_{\rm
met})C_{\rm CO}$. These calculated $C_{\rm tp}(T)$ curves for
$y=0.4$, 0.5, and 0.6 samples are shown in Fig. 3a as dashed
lines. These curves show a reasonable quantitative agreement with
the experimental data, which argues strongly for the coexistence
of CO and metallic phases at low temperatures in these materials.
However, this approach does not give an exact agreement with the
experimentally observed specific heat. A fit of the data to the
form of Eq. (2) gives  much better agreement with the data as
shown by the solid lines in Fig. 3b. The values of the fitting
parameters are shown in Table I, where we consider $C^{\prime}(T)$
to be a contribution from nonmagnetic excitations with dispersion
relation $\epsilon = \Delta + Bq^2$ ($\Delta $ is an energy gap
and $q$ is a wave vector) \cite{Vera2,Vera1}. These results show
that the specific heat of $y=0.4$, 0.5, and 0.6 samples is not
just the linear combination of $C_{\rm met}$ and $C_{\rm CO}$.
Since the phase separation in
(La$_{1-y}$Pr$_y$)$_{0.67}$Ca$_{0.33}$MnO$_{3}$ is electronic (not
chemical), the presence of both La and Pr ions could modify the
metallic and CO phase in these materials, and CO and metallic
phases in phase separated regions of
(La$_{1-y}$Pr$_y$)$_{0.67}$Ca$_{0.33}$MnO$_{3}$  differ from CO
and metallic phases of Pr$_{0.65}$Ca$_{0.35}$MnO$_{3}$ and
La$_{0.7}$Ca$_{0.3}$MnO$_{3}$ respectively. The following analysis
of the specific heat of
(La$_{1-y}$Pr$_y$)$_{0.67}$Ca$_{0.33}$MnO$_{3}$  samples will be
done using Eq.(2), since the volume fraction approach gives less
close agreement with the data.


\subsection{Specific heat, resistivity, and magnetization of
 (La$_{1-y}$Pr$_y$)$_{0.67}$Ca$_{0.33}$MnO$_{3}$ with $^{16}$O and $^{18}$O}

\subsubsection{Low temperature specific heat and resistivity}

The low temperature specific heat and resistivity of $y=0.4$ and
0.5 samples with $^{16}$O and $^{18}$O is shown in Fig. 4. The
resistivity  of these compositions is affected significantly by
the oxygen isotope substitution (Fig. 4, insets): the onset of
metallic behavior (the peak temperature for cooling run) decreases
by 50 K for the $y = 0.5$ sample and by 28 K for the $y = 0.4$
sample; the low temperature value of the resistivity increases
approximately by one order in magnitude for both compositions. The
\begin{figure}[tbp]
\centerline{
\psfig{figure=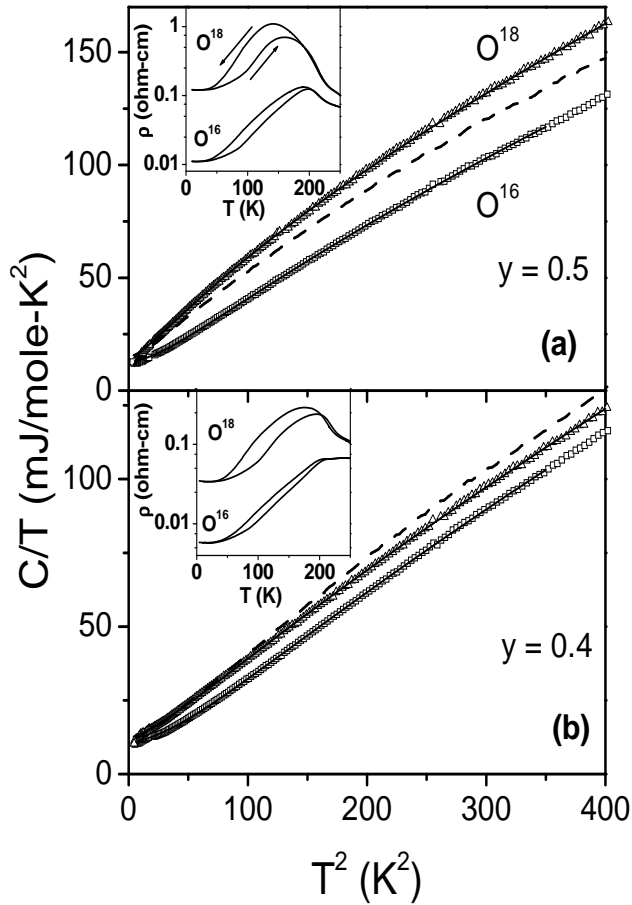,width=8.7cm,height=12.5cm,clip=} }
\caption{Specific heat of
(La$_{1-y}$Pr$_y$)$_{0.67}$Ca$_{0.33}$MnO$_{3}$ with $^{16}$O
(squares) and $^{18}$O (triangles): (a) $y=0.5$; (b) $y = 0.4$.
Dashed line is specific heat of $y=0.6$ sample in (a) and $x=0.5$
sample in (b). Insets show the temperature dependence of the
resistivity for $^{16}$O and $^{18}$O samples on warming and
cooling. Solid lines are fits described in text. } \label{fig4}
\end{figure}
\noindent low temperature resistivity of metallic
La$_{1-x}$Ca$_{x}$MnO$_{3}$ did not show such dramatic effects
\cite{GMltR}. This behavior is consistent with previous work:
substitution  of  $^{16}$O by $^{18}$O  favors the insulating
state \cite{TCO,MI}.

Unlike in the single phase metallic and CO compositions of
La$_{1-x}$Ca$_{x}$MnO$_{3}$ (Fig. 1), in phase separated
(La$_{1-y}$Pr$_y$)$_{0.67}$Ca$_{0.33}$MnO$_{3}$ the substitution
of $^{16}$O by $^{18}$O induces substantial changes in the low
temperature specific heat (Fig. 4). The specific heat, dominated
by the $C^{\prime}(T)$ contribution in most of the temperature
range, is larger for the $^{18}$O samples. This increase of the
specific heat for the $^{18}$O  sample is similar to the increase
of the specific heat with an increase of the Pr content $y$.  The
specific heat of the $y=0.6$ sample is shown in Fig. 4a for
comparison. A result of the fits of our data to Eq. (2) is given
in Table I. All fitting parameters are close to those of the
higher $y$ value samples. These results indicate that the amount
of the CO phase is larger in $^{18}$O  samples.

\subsubsection{High temperature specific heat and magnetization}

The specific heat of $y=0.5$ $^{16}$O and $^{18}$O samples in the
temperature range from 2 K to 270 K is shown in Fig. 5a. The room
temperature value of the specific heat, 110 J/mole K, is in
agreement with the room temperature values  reported previously in
manganites \cite{sph}. For both samples we observed a broad
anomaly associated with transition to a magnetically and charge
ordered state at $T \approx 220$ K. Similar behavior was reported
for an $^{16}$O sample of similar composition \cite{Kiryukhin},
where the anomaly in C(T) was attributed to the CO detected by
x-ray diffraction. The temperature dependence of the magnetization
measured in $\mu _0 H =50$~G and the inverse susceptibility is
shown in the inset of Fig. 5a. The onset of the magnetic
transition at $\approx 220$ K in both samples is best seen from
the temperature dependence of the inverse susceptibility. The high
temperature inverse susceptibility of both samples (which
extrapolates to intersect the x-axis at positive value) indicates
a FM interaction. However, for the $^{18}$O sample, the value of
the inverse susceptibility increases (the magnetization drops)
just below the transition at 220 K, which is indicative of a
transition to the AFM state. This suggests a presence of competing
FM and AFM interactions leading to a two phase coexistence in
these materials. Ferromagnetic ordering in the $^{18}$O sample is
established below $\approx 100$ K. The $^{16}$O sample exhibits a
higher value of the magnetization below 220 K than expected for
the AFM state, perhaps due to the presence of a fraction of the FM
phase. The magnetization    of the $^{16}$O sample increases
further below 120 K indicating that a larger volume fraction
becomes FM. These magnetization data of the $^{16}$O sample are
consistent with \cite{Cheong1}, where the transition to a FM state
from the CO AFM state and corresponding insulator to metal
transition was shown to be of a percolative nature. The
substitution of $^{16}$O by $^{18}$O does not change the
temperature of the magnetic and CO transitions in this material
but decreases the volume fraction of the FM phase. The specific
heat does not show an anomaly associated with the metal-insulator
transition for both samples, which is consistent with percolative
nature of the transition occurring gradually over a wide
temperature range.

\begin{figure}[tbp]
\centerline{
\psfig{figure=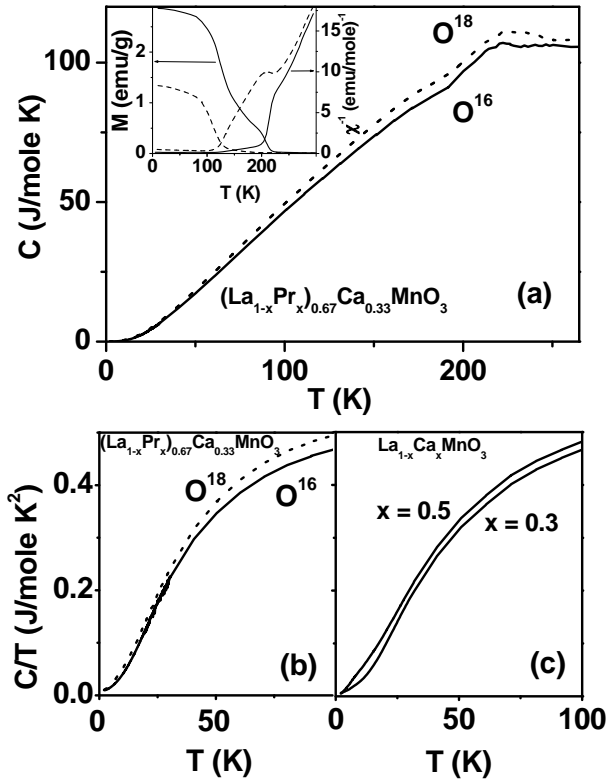,width=8.7cm,height=11.5cm,clip=} }
\caption{Specific heat of
(La$_{1-y}$Pr$_y$)$_{0.67}$Ca$_{0.33}$MnO$_{3}$ $y=0.5$ with
$^{18}$O (dashed line) and $^{16}$O (solid line) in the
temperature interval: (a) from 2 K to 300 K; (b) enlarged region
from 2 K to 100 K. Inset shows the temperature dependence of the
magnetization and inverse susceptibility for these samples. (c)
Specific heat of CO La$_{0.5}$Ca$_{0.5}$MnO$_{3}$ and metallic
La$_{0.7}$Ca$_{0.3}$MnO$_{3}$. } \label{fig5}
\end{figure}

The low temperature specific heat (2 K$ < T < $19 K) of the $^{18}$O sample is
higher than that of $^{16}$O sample (Fig. 4a). This tendency persists also
at higher temperatures (see Fig. 5a,b). This difference in specific heat of $^{18}$O and
$^{16}$O samples is not primarily caused by an enhanced lattice contribution from the heavier
atom: the difference in the lattice contributions at $T=70$ K, estimated from
the Debye function, should be $\approx 1$\%, while the observed difference
is $\approx 6$\%. A similar behavior is found for La$_{0.5}$Ca$_{0.5}$MnO$_{3}$,
which has an excess specific heat  compared with metallic
La$_{0.7}$Ca$_{0.3}$MnO$_{3}$ (see Fig. 5c). This implies that the excitations responsible
for the excess specific heat in the CO state \cite{Vera2} are also contributing
at higher temperatures. In the case of
(La$_{1-y}$Pr$_y$)$_{0.67}$Ca$_{0.33}$MnO$_{3}$ ($x=0.5$) samples,
the larger specific heat of the $^{18}$O sample (which comes from the
excitations in the CO states) indicates a larger volume
fraction of the CO phase over a wide temperature range.

\subsubsection{Effect of a magnetic field}

The field dependence of the magnetization for $y=0.5$ and 0.4
$^{16}$O and $^{18}$O samples at $T=5.5$ K is shown in Fig. 6.
Samples were cooled in zero magnetic field (ZFC) and the
hysteresis loops in a magnetic field up to 9 T were taken. After
ZFC the samples appear to be in a state with smaller than full
spin alignment magnetization value. This indicates the FM and AFM
phase coexistence in these materials. The values of the ZFC
magnetization at $\mu _0 H = 1$ T (where the saturation for the FM
phase is reached) are smaller for the materials with $^{18}$O
which corresponds to the larger fraction of the AFM phase in the
$^{18}$O samples. At $\mu _0 H  \approx 1.5$ T the magnetic field
starts to affect the CO (and associated AFM ordering) which causes
an increase of the magnetization. In the $^{18}$O samples the
transition is completed in magnetic fields higher than in $^{16}$O
samples. Even when the transition is completed ($\mu _0 H >8$~T),
the magnetization value of the $^{18}$O samples is somewhat
smaller than that of the $^{16}$O samples. These magnetization
measurements show that substitution of $^{16}$O by $^{18}$O not
only increases the fraction of the AFM CO phase, as shown by
specific heat, but also lowers its energy, since it requires a
higher magnetic field to stabilize the FM state  than for the
$^{16}$O samples. As in other CO compositions, once the FM phase
is stabilized, the material remains in FM state even though the
magnetic field is removed, and only heating the sample above the
CO transition temperature can return it to the state with lower
magnetization value.

\begin{figure}[tbp]
\centerline{
\psfig{figure=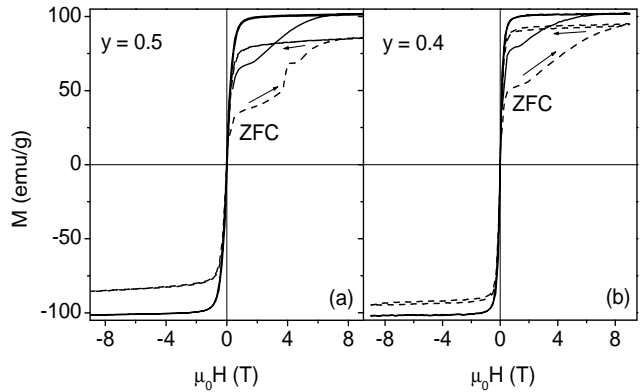,width=8.7cm,height=5.5cm,clip=} }
\caption{The magnetic field dependence of the magnetization of the
(La$_{1-y}$Pr$_y$)$_{0.67}$Ca$_{0.33}$MnO$_{3}$ with $^{16}$O
(solid lines) and $^{18}$O (dashed lines) samples for (a) $y=0.5$
and (b) 0.4 compositions at $T = 5.5$ K. } \label{fig6}
\end{figure}

Application of magnetic field of 8.5~T reduces the resistivity of the
(La$_{1-y}$Pr$_y$)$_{0.67}$Ca$_{0.33}$MnO$_{3}$ $y=0.5$ $^{16}$O and $^{18}$O
samples (Fig. 7, insets). The resistivity of both samples in $\mu _0 H  = 8.5$ T has a
metallic temperature dependence (for $T<300$ K) with values essentially the
same for $^{16}$O and $^{18}$O and close to the resistivity of the metallic
La$_{0.7}$Ca$_{0.3}$MnO$_{3}$.

\begin{figure}[tbp]
\centerline{
\psfig{figure=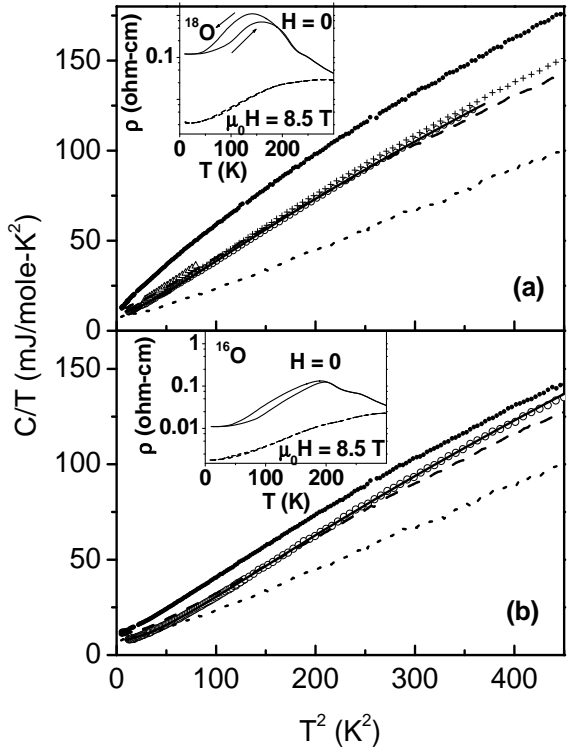,width=8.5cm,height=11.5cm,clip=} }
\caption{Specific heat of
(La$_{1-y}$Pr$_y$)$_{0.67}$Ca$_{0.33}$MnO$_{3}$ $y=0.5$ samples.
(a) $^{18}$O: filled circles - $\mu _0 H =0$, open circles - $\mu
_0 H =8.5$ T, solid line - fit to the 8.5 T data, crosses - $\mu
_0 H =4$ T (field was reduced to 4 T after the application of 8.5
T), triangles $H=0$ (also after 8.5 T); dashed line - $y=0.5$,
$^{16}$O, $\mu _0 H =0$; dotted line - $y=0$, $^{16}$O, $H=0$
(shown for comparison). (b) $^{16}$O: filled circles - $H=0$, open
circles - $\mu _0 H =8.5$ T, solid line - fit to the 8.5 T data;
dashed line - $y=0.4$, $^{16}$O, $H=0$; dotted line - $y=0$,
$^{16}$O, $\mu _0 H =0$ (shown for comparison). Insets show the
temperature dependence of the resistivity in $H=0$ and $\mu _0 H
=8.5$ T for $y=0.5$ samples: (a) $^{18}$O, (b) $^{16}$O. }
\label{fig7}
\end{figure}

In $\mu _0 H =8.5$ T the specific heat of
both samples drops significantly (Fig. 7, open circles), and tends to retain
this low value  when the magnetic field is reduced to 4 T
(Fig.7a, crosses) and 0 T (triangles). This irreversible behavior is clearly
demonstrated in the $C(H)$ plot (Fig. 8). This is similar to the behavior of
the magnetization -
only by heating the sample above the CO transition temperature does
the specific heat return to its higher value. We reported similar magnetic
and thermal history
dependencies associated with ``melting" of the CO in magnetic field
for Pr$_{0.65}$Ca$_{0.35}$MnO$_{3}$ \cite{Vera1}. As in
Pr$_{0.65}$Ca$_{0.35}$MnO$_{3}$, the specific heat at $\mu _0 H  = 8.5$ T
(magnetic field sufficient to
induce a transition to the FM state (Fig. 6))
is significantly larger than $C(T)$ of metallic La$_{0.7}$Ca$_{0.3}$MnO$_{3}$
(dotted line in Figs. 7a and b) and exhibits the characteristic
upward curvature of the CO state in the $C/T$ vs. $T^2$ plot.
The specific heat  in 8.5 T is similar to the specific heat of  samples
with a smaller fraction of the CO: for the $^{18}$O sample, $C(T)$ is close to the zero
field $C(T)$ of $^{16}$O
$y=0.5$ (dashed line in Fig. 7a), and for the $^{16}$O sample $C(T)$ is close to the zero
field $C(T)$ of
$y=0.4$ (dashed line in Fig. 7a).  The results of fitting the 8.5 T data to Eq. 2
(see Table I)
gives values close to the fitting parameters of the samples
with the smaller fraction of the CO.

\begin{figure}[tbp]
\centerline{
\psfig{figure=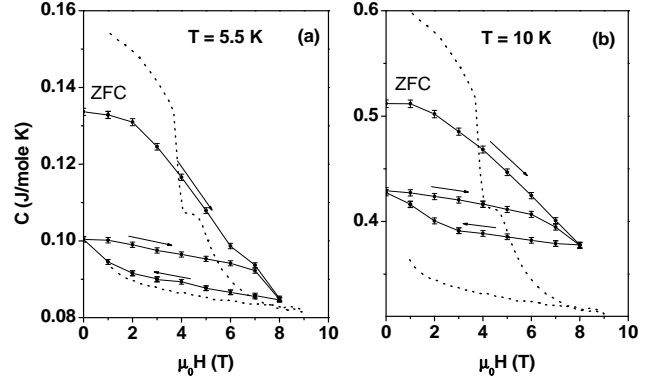,width=8.7cm,height=5.5cm,clip=} }
\caption{Specific heat of the
(La$_{1-y}$Pr$_y$)$_{0.67}$Ca$_{0.33}$MnO$_{3}$ $y=0.5$ sample
with $^{18}$O as a function of magnetic field at: (a) $T=5.5$ K
and (b) $T=10$ K. Solid lines are a guide to the eye. Dotted lines
are calculated specific heat described in text.  } \label{fig8}
\end{figure}

Unlike the magnetization, the $C(H)$ (when the field is applied to
a ZFC sample) at $T = 5.5$ K and 10 K shows no saturation in a
magnetic field of 8 T (Fig.8). If we attempt to relate the value
of the magnetization to the fraction of the metallic phase and the
rest to the fraction of the CO phase, as we did above for
different $y$, we can calculate the field dependence of the
specific heat as $C_{\rm tp}(H)=f_{\rm met}C_{\rm met}+(1-f_{\rm
met})C_{\rm CO}$. This calculated specific heat of the two phase
mixture $C_{\rm tp}(H)$ is not in agreement with the experimental
data (Fig. 8): the ZFC calculated $C_{\rm tp}(H)$ (dotted lines in
Fig. 8) decreases rapidly in the field interval from 4 T to 6 T
and saturates at $\mu _0 H =7$ T, while the ZFC experimental
$C(H)$ data have approximately the same slope up to $\mu _0 H =8$
T without a sign of saturation. As we noted before, the
magnetization does not recover its full FM value even in magnetic
field of 9 T, which probably means that some fraction of the AFM
phase still remains at this field. These results indicate that the
CO is not completely destroyed by the magnetic field sufficient to
induce the FM transition. This phenomenon appears to be  common
for all CO manganites.

The magnetic field has a different effect on the samples with different oxygen isotopes.
$C(T)$ is higher for the $^{18}$O sample due to larger fraction of the CO phase
remaining in this magnetic field, since  $C(T)$  (Fig. 7a) is close to
to the zero field $C(T)$ of the $^{16}$O with $y=0.5$, while the $C(T)$ of the $^{16}$O sample
in 8.5 T is close to $C(T)$ of the $y=0.4$ sample which has smaller fraction
of the CO phase. This has qualitative  agreement with the magnetization: larger
magnetic field is needed to be applied to the $^{18}$O samples to induce the same
effect on the charge ordering as for the $^{16}$O samples.

\section{DISCUSSION}

Our study shows that the principal excitations of single phase
systems such as metallic compositions of
La$_{1-x}$Ca$_{x}$MnO$_{3}$ and insulating charge-ordered
La$_{0.5}$Ca$_{0.5}$MnO$_{3}$ are not affected by oxygen isotope
exchange. A recent theoretical model \cite{Alexandrov}, which
suggests that the ground state of metallic manganites is a
polaronic Fermi-liquid, predicts an increase of the effective mass
(and hence $\gamma$) by about 3-6 \% for the $^{18}$O material.
Other authors point out that a large oxygen isotope effect on the
effective mass of the charge carriers can only be expected for
systems close to a local structural instability \cite{Millis}.
Recent neutron scattering experiments have shown that the
spin-wave dynamics and therefore the exchange interaction is the
same  for ferromagnetic La$_{1-x}$Ca$_{x}$MnO$_{3}$  with $^{16}$O
and $^{18}$O \cite{Lynn}. Also,  Mossbauer effect measurements
have shown that the exchange interaction does not change with
oxygen isotope substitution \cite{Nath}. These results also show
that the metallic ground state of these materials is not affected
by the oxygen isotope exchange, which is in agreement with our
specific heat data because of the following reasons. Within the
double exchange model, the exchange interaction is determined by
the an effective transfer integral $t_{\rm eff}$ for an electron
hopping between Mn ions. In the tight-binding approximation, the
density of states at the Fermi level, $N(E_{\rm F})$, is inversely
proportional to $t_{\rm eff}$. Since the $N(E_{\rm F})$ is
directly proportional to $\gamma$, materials with the same
exchange interaction should have the same $\gamma$. In contrast,
drastic changes in the low temperature properties are observed for
the phase separated
(La$_{1-y}$Pr$_y$)$_{0.67}$Ca$_{0.33}$MnO$_{3}$, associated, as we
have shown, with an increase of the CO phase fraction in the
$^{18}$O materials.

A large isotope effect was observed for the ferromagnetic transition
in metallic compositions of La$_{1-x}$Ca$_{x}$MnO$_{3}$, but only a
negligible effect was found for
La$_{1-x}$Sr$_{x}$MnO$_{3}$ \cite{GMNature}. The charge ordering temperature
changes significantly with isotope exchange for La$_{0.5}$Ca$_{0.5}$MnO$_{3}$ and Nd$_{0.5}$Sr$_{0.5}$MnO$_{3}$,
but not for Pr$_{0.5}$Ca$_{0.5}$MnO$_{3}$
\cite{TCO}. So what is in common for the large isotope effect materials?
We believe it is the fact that the phase transition is first order and
a phase separation is observed in  the temperature region where the transition occurs.
 The FM transition in La$_{1-x}$Ca$_{x}$MnO$_{3}$
is believed to be the first order because the heat capacity peak
associated with FM transition shifts significantly with
application of magnetic field \cite{CIllinois} and the spin
dynamics observed in the vicinity of the FM transition is not
consistent with a second order transition \cite{LynnD}. Moreover,
there is strong evidence for  phase separation in the vicinity of
the Curie temperatures for La$_{1-x}$Ca$_{x}$MnO$_{3}$
\cite{fath,FMps}. In contrast, La$_{1-x}$Sr$_{x}$MnO$_{3}$ does
not exhibit a phase separation near the FM transition and the
phase transition in this material is second order \cite{Kartik}.
The charge ordering transition for La$_{0.5}$Ca$_{0.5}$MnO$_{3}$
and Nd$_{0.5}$Sr$_{0.5}$MnO$_{3}$ is a first order transition from
a FM metal phase to a AFM CO phase accompanied by a hysteresis,
showing the presence of both phases in the transition region. But
Pr$_{0.5}$Ca$_{0.5}$MnO$_{3}$ has a transition from a charge
disordered to a charge ordered phase without signs of phase
separation \cite{Mori} and this transition is probably second
order.

As was pointed out by some authors \cite{Millis,Khomskii}, the polaron effect
alone \cite{GMNature}
is too small to  induce the observed changes associated with oxygen isotope exchange.
Perhaps the polaron bandnarrowing effect produced by $^{18}$O induces localization effects
in the phase separated region, which stabilizes the CO phase and leads to the increase of
the volume fraction of the CO phase, as we observe for
(La$_{1-y}$Pr$_y$)$_{0.67}$Ca$_{0.33}$MnO$_{3}$. This leads to a decrease of the Curie
temperature and an increase of the CO temperature in the phase separated system. In this picture,
the largest effects should be observed near the percolation threshold, which is consistent
with our observation that (La$_{1-y}$Pr$_y$)$_{0.67}$Ca$_{0.33}$MnO$_{3}$ with
$y=0.5$ exhibit
larger changes with oxygen isotope exchange than $y=0.4$, and with the fact that for
$y=0.75$ the oxygen isotope exchange induces a metal to insulator transition \cite{MI}.

The experimental facts discussed above suggest: 1) the free energy
of the CO phase is lower for $^{18}$O than for $^{16}$O, 2) the
free energy of the metallic phase does not change much with oxygen
isotope exchange,  and 3) the large isotope effect on the
transition temperature occurs when the transition is of first
order (phase separation occurs). Fig. 9a shows a schematic free
energy diagram in the vicinity of the FM - CO transition which
occurs for La$_{0.5}$Ca$_{0.5}$MnO$_{3}$ and
Nd$_{0.5}$Sr$_{0.5}$MnO$_{3}$ compositions. Since the transition
is first order, the free energy curve $F(T)$  of the CO state
crosses the $F(T)$  of the FM state at $T_{\rm CO}$ and the CO
state has a lower value of $F(T)$ at $T<T_{\rm CO}$. The $F(T)$ of
the CO phase of $^{18}$O material is lower than the $^{16}$O
material. Therefore, the $F(T)$  of the CO phase of $^{18}$O
material intersects the $F(T)$ of the FM phase at higher
temperature, $T_{\rm CO}$($^{18}$O) $>$ $T_{\rm CO}$($^{16}$O).
Using this picture, the change in the CO temperature $\Delta
T_{\rm CO}$ = $T_{\rm CO}$($^{18}$O) - $T_{\rm CO}$($^{16}$O) can
be approximated as $\Delta T_{\rm CO}$ = $\delta / \Delta S$,
where $\delta $ is the difference between the free energy of the
$^{16}$O and $^{18}$O  CO phases, and $\Delta S$ is the difference
in entropy (d$F$/d$T$) of the CO and FM phases at the transition
temperature.

The entropy change associated with this transition was found to be
5~J/mole~K for La$_{0.5}$Ca$_{0.5}$MnO$_{3}$ \cite{Ramirez}. To
estimate $\delta$, we use the experimental fact that the magnetic
field sufficient to destroy the CO state and induce a transition
to the FM state is higher for $^{18}$O than for $^{16}$O (see, for
example, Fig. 6). The gain in magnetic energy associated with the
transition to the FM state is $MH_{\rm FM}$, where $M$ is the
magnetization of the FM state. The difference between the $MH$
values for $^{18}$O and $^{16}$O materials should correspond to
the difference in free energy of the CO state of the $^{18}$O and
$^{16}$O materials. The transition to the FM state for
La$_{0.5}$Ca$_{0.5}$MnO$_{3}$ occurs at magnetic fields higher
than 9 T \cite{Xiao}, beyond our experimental field limit.
Therefore, we approximate the difference in free energy of the CO
state of the $^{18}$O and $^{16}$O materials from our $M(H)$ data
for (La$_{1-y}$Pr$_y$)$_{0.67}$Ca$_{0.33}$MnO$_{3}$  (Fig.~6).
Although the transition to the FM state is very broad for both
$^{18}$O and $^{16}$O samples, a saturation occurs for the
$^{18}$O sample at a magnetic field 2 T higher than for the
$^{16}$O sample. Thus, $\delta \approx MH_{\rm FM}$($^{18}$O) -
$MH_{\rm FM}$($^{16}$O) $\approx 42$ J/mole, and estimated
increase in the CO transition temperature $\Delta T_{\rm CO} =
\delta / \Delta S  \approx 8.6$ K, which is the same order as the
observed  $\Delta T_{\rm CO} = 11.7$ K. This agreement suggests
that the proposed free energy diagram (Fig. 9a) is valid.
\begin{figure}[tbp]
\centerline{
\psfig{figure=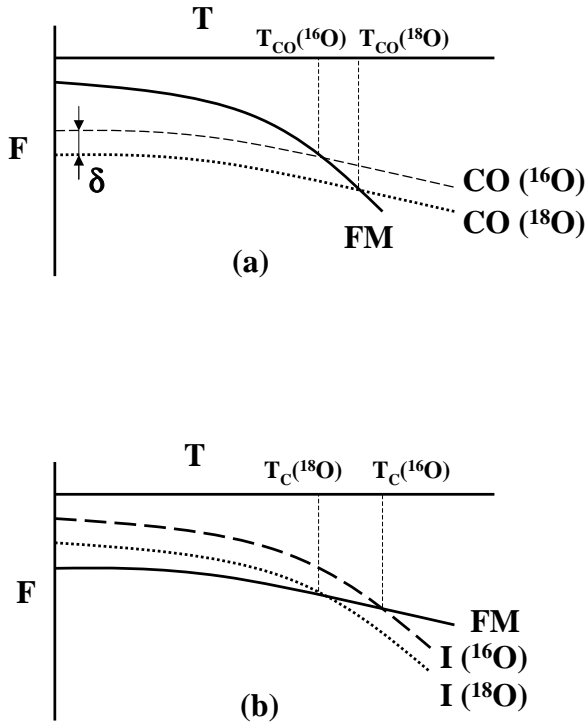,width=8.5cm,height=10.5cm,clip=} }
\caption{(a) Schematic free energy diagram in the vicinity of the
FM - CO transition. Solid line is the free energy, $F(T)$, of the
FM state, dashed line is the $F(T)$ of the $^{16}$O charge-ordered
phase, dotted line is the $F(T)$ of the $^{18}$O CO phase. (b)
Schematic free energy diagram in the vicinity of the FM
transition. Solid line is the $F(T)$ of the FM state, dashed line
is the $F(T)$ of the $^{16}$O insulating phase, dotted line is the
$F(T)$ of the $^{18}$O insulating phase.} \label{fig9}
\end{figure}
In the case of the FM transition in La$_{1-x}$Ca$_{x}$MnO$_{3}$
($0.2<x<0.4$) in the vicinity of the FM transition the free energy
curve $F(T)$  of the insulating state crosses the $F(T)$  of the
FM state at $T_{\rm C}$ and the FM state has a lower value of
$F(T)$  at $T<T_{\rm C}$ (Fig.9b). The insulating state of
La$_{1-x}$Ca$_{x}$MnO$_{3}$ ($0.2<x<0.4$) shows a signature of
short range charge ordering \cite{FMps} in the vicinity of the FM
transition. We argue that the substitution of $^{18}$O for
$^{16}$O increases the volume fraction of the CO phase in the
vicinity of the FM transition, which lowers the free energy of the
insulating state. This shifts the crossing of $F(T)$  of the
insulating state with $F(T)$  of the FM state to lower temperature
(Fig. 9b), in agreement with the lowering of $T_{\rm C}$ observed
when $^{18}$O replaces $^{16}$O in these materials.

\section{CONCLUSIONS}

We have studied the influence of oxygen isotope exchange on the
resistivity, magnetization, and specific heat of single phase and
phase separated manganites. We have found that the principal
excitations in single phase metallic and CO materials do not
change with oxygen isotope substitution, while isotope exchange
induces large changes in the low temperature properties of phase
separated (La$_{1-y}$Pr$_y$)$_{0.67}$Ca$_{0.33}$MnO$_{3}$.
The variation of the specific heat with Pr content
$y$ in (La$_{1-y}$Pr$_y$)$_{0.67}$Ca$_{0.33}$MnO$_{3}$ samples is
consistent with the larger volume fraction of the CO phase for higher $y$.
Oxygen isotope exchange in these materials induced changes similar
to the increase of the Pr content $y$ which suggests an increase of the volume
fraction of the CO phase in $^{18}$O samples.

Similar to other CO materials (Pr$_{1-y}$Ca$_{y}$MnO$_{3}$,
La$_{0.5}$Ca$_{0.5}$MnO$_{3}$) all studied composition of
(La$_{1-y}$Pr$_y$)$_{0.67}$Ca$_{0.33}$MnO$_{3}$ ($0.4<y<1$) have
shown the presence of the anomalous $C'(T)$ contribution to the
low temperature specific heat due to the presence  of the CO
phase. The excess specific heat is also present at higher
temperatures ($T<T_{\rm CO}$) for the CO materials, or materials
containing larger fraction of the CO phase. Specific heat
measurements indicate that the charge ordering is not destroyed
completely by the application of a magnetic field of 8.5 T in
contrast to the resistivity which shows a compleat ``melting" of
the CO. The nature  of the excitations responsible for the $C'(T)$
term in the specific heat of the CO materials, and the presence of
this term  in rather strong magnetic field, has yet to be
explained.

We suggest that the giant isotope effect on the Curie temperature and
the CO temperature in manganites occurs in materials where these
transitions are first order and hence, where the CO phase is present in
the vicinity
of the phase transition. $^{18}$O  lowers the energy of the CO phase,
which leads
to an increase in the volume fraction of the CO phase in the phase
separated regions.
This leads to a decrease of the Curie temperature and increase of
the CO temperature. A detailed understanding of the mechanism  lowering
the energy of the CO phase in the $^{18}$O  materials still  needs to be developed.

\noindent {\bf Acknowledgment}: We thank A. J. Millis for helpful
discussions, S-W. Cheong for providing us with as-grown
La$_{0.625}$Ca$_{0.375}$MnO$_{3}$ sample and helpful discussion,
Z. Li for the sample preparation. The work at Maryland is
supported in part by the NSF-MRSEC, DMR \#00-80008. The work in
Sherbrooke is supported by the Canadian Institute of Advanced
Research (CIAR), the Canadian Foundation for Innovation (CFI), the
Natural Sciences and Engineering Research Council of Canada
(NSERC) and the Fondation FORCE of the Universite de Sherbrooke.

\end{document}